\documentclass[useAMS,usegraphicx]{mn2e}

\newcommand{\Msun}{\ensuremath{\,{\rm M}_\odot}}            
\newcommand{\Rsun}{\ensuremath{\,{\rm R}_\odot}}            
\newcommand{\Lsun}{\ensuremath{\,{\rm L}_\odot}}            
\newcommand{\logg}{\ensuremath{\log g}}                     
\newcommand{\Vrot}{\ensuremath{V_{\rm rot}}}                
\newcommand{\EBV}{\ensuremath{E_{B-V}}}                     
\newcommand{\Eby}{\ensuremath{E_{b-y}}}                     
\newcommand{\kms}{\,km\,s$^{-1}$}                           
\newcommand{\ion}[2]{{#1}\,{\sc {\small{#2}}}}              
\newcommand{\etal}{\mbox{et~al.}}

\title[Eclipsing binaries in open clusters. I. V615\,Per and V618\,Per in h\,Persei]
      {Eclipsing binaries in open clusters. \\ I. V615\,Per and V618\,Per in h\,Persei 
       \thanks{Based on observations made with the INT and JKT operated on the island of La Palma by the Isaac Newton 
        Group in the Spanish Observatorio del Roque de los Muchachos of the Instituto de Astrofis\'\i ca de Canarias}}

\author[J.\ Southworth, P.\ F.\ L.\ Maxted and B.\ Smalley]
       {J.\ Southworth$^1$\thanks{Email addresses: jkt,pflm,bs@astro.keele.ac.uk}, 
        P.\ F.\ L.\ Maxted$^1$\footnotemark[2] and B.\ Smalley$^1$\footnotemark[2] \\
        $^1$\,Department of Physics, Keele University, Staffordshire, ST5 5BG, UK}

\begin{document} \maketitle 

\begin{abstract}
We derive absolute dimensions for two early-type main sequence detached eclipsing binaries in the young open cluster h\,Persei (NGC\,869). V615\,Persei has a spectral type of B7\,V and a period of 13.7 days. V618\,Persei is A2\,V and has a period of 6.4 days. New ephemerides are calculated for both systems. 
The masses of the component stars have been derived using high-resolution spectroscopy and are $4.08 \pm 0.06$\Msun\ and $3.18 \pm 0.05$\Msun\ for V615\,Per and $2.33 \pm 0.03$\Msun\ and $1.56 \pm 0.02$\Msun\ for V618\,Per. The radii have been measured by fitting the available light curves using {\sc ebop} and are $2.29 \pm 0.14$\Rsun\ and $1.90 \pm 0.09$\Rsun\ for V615\,Per and $1.64 \pm 0.07$\Rsun\ and $1.32 \pm 0.07$\Rsun\ for V618\,Per.
By comparing the observed spectra of V615\,Per to synthetic spectra from model atmospheres we find that the effective temperatures of the two stars are $15000 \pm 500$\,K and $11000 \pm 500$\,K. The equatorial rotational velocities of the primary and secondary components of V615\,Per are $28 \pm 5$\kms\ and $8 \pm 5$\kms, respectively. Both components of V618\,Per rotate at $10 \pm 5$\kms. The equatorial rotational velocities for synchronous rotation are about 10\kms\ for all four stars.
The timescales for orbital circularisation for both systems, and the timescale for rotational synchronisation of V615\,Per, are much greater than the age of h\,Per. Their negligible eccentricities and equatorial rotational velocities therefore support the hypothesis that they were formed by `delayed breakup' (Tohline 2002).
We have compared the radii of these stars to models by the Granada and the Padova groups for stars of the same masses but different compositions. We conclude that the metallicity of the stars is $Z \approx 0.01$. This appears to be the first estimate of the bulk metallicity of h\,Per. Recent photometric studies have assumed a solar metallicity so their results should be reviewed.
\end{abstract}



\section{Introduction}        \label{introduction}    

Current theoretical stellar models are increasingly sophisticated and accurate, and different sets of models often produce very similar predictions for the physical parameters of stars. This is particularly pronounced on the main sequence, where evolution is slow and agreement between models is very good (Pols \etal\ 1997). Observational tests of stellar models therefore require high-quality data and careful treatment to allow a discriminating test of the success of a particular set of physical ingredients compared to other evolutionary models (Andersen 1991). This is exacerbated by the number of physical parameters which are not observed or well known, so several quantities can be freely adjusted when attempting to fit observed data, e.g., the abundances of helium and metals, the degree of convective overshooting (Young \etal\ 2001) and indeed the type of convective theory.

Detached eclipsing binaries (dEBs) have often been used for testing theoretical models. The combination of a double-lined spectroscopic orbit and the fitting of a simple geometrical model to an observed light curve allows the derivation of accurate masses, radii, and radiative properties of the two stars (Andersen 1991). The component stars are expected to have the same age and chemical composition, but models can still be chosen of any age and composition (within reason) to fit the observed stellar parameters.

Photometric studies of open clusters are a very common method of testing the predictions of stellar models (e.g., Daniel \etal\ 1994). This method has the potential to test subtle effects and might require little telescope time, but it can be limited in its usefulness due to degeneracies between the distance to the cluster, its age, metallicity and reddening effects (including differential reddening over the cluster). It can also be difficult to identify unresolved binary stars and field stars, which can affect the reliability of the conclusions (Nordstr\"om, Andersen \& Andersen 1997). A further problem is the need to convert observed quantities to theoretical values to properly compare observations to stellar models. 

\begin{table} \begin{center} 
\caption{\label{photpartable} Identifications and combined photometric indices for V615\,Per and V618\,Per from various studies. The more recent Str\"omgren photometry of Capilla \& Fabregat (2002) is not preferred because the data for V615\,Per suggest it was in eclipse during some of their observations. All photometric parameters (including the spectral type determined from the Str\"omgren colours) refer to the combined system light. 
\newline $^*$\,Calculated from the system magnitude in the $V$ filter, the adopted cluster distance modulus and reddening (see section~\ref{clusterinfo}) and the canonical reddening law $A_V = 3.1 \EBV$.
\newline {\bf References:} (1) Oosterhoff (1937); (2) Keller \etal\ (2001); (3) Slesnick \etal\ (2002); (4) Capilla \& Fabregat (2002); (5) Marco \& Bernabeu (2001); (6) Uribe \etal\ (2002) based on proper motion and position.} 
\begin{tabular}{lr@{}lr@{}lr} \hline \hline
\                             &   & V615\,Per       &   & V618\,Per       & Ref       \\ \hline
Oosterhoff number             &   & Oo1021          &   & Oo1147          & 1         \\
Keller number                 &   & KGM 644         &   & KGM 1901        & 2         \\
Slesnick number               &   & SHM 663         &   & SHM 1965        & 3         \\ \hline
$\alpha_{2000}$               &   & 2 19 01.65      &   & 2 19 11.85      & 4         \\
$\delta_{2000}$               & + & 57 07 19.2      & + & 57 06 41.2      & 4         \\ \hline
$V$                           &   & 13.015          &   & 14.621          & 3         \\
$B-V$                         &   & 0.388           &   & 0.613           & 3         \\
$U-B$                         &$-$&0.101            &   & 0.286           & 3         \\
$V-I$                         &   & 0.370           &   & 0.661           & 2         \\ \hline
$b-y$                         &   & 0.351           &   & 0.473           & 5         \\
$m_1$                         &$-$&0.020            &   & 0.010           & 5         \\
$c_1$                         &   & 0.601           &   & 0.998           & 5         \\
$\beta$                       &   & 2.762           &   & 2.861           & 5         \\
Photo.\ spectral type         &   & B8              &   & A3              & 5         \\ 
$M_V$                         &$-$&0.45             &   & 1.15            & $^*$      \\ \hline
Membership prob.              &   & 0.96            &   & --              & 6         \\
\hline \hline \end{tabular} \end{center} \end{table}

Eclipsing binaries in open clusters provide a way of combining these two methods to reduce the number of free parameters which can be adjusted until the models provide a good fit. The two datasets are complementary and can be used in two ways. If the primary interest is in dEBs then the masses and radii of two stars are supplemented by some knowledge of the age and chemical composition of the system from its membership of an open cluster. Alternatively, a normal study of the positions of stars in photometric diagrams can be improved by anchoring the stellar models used to the observed masses, radii, and effective temperatures for the components of the dEB. 

A further advantage of having accurate radii of stars in a dEB is that the distance to each star can be found independently of stellar models. The usual method is to calculate bolometric luminosities from stellar effective temperatures, radii and bolometric corrections, but this suffers from the low accuracy of empirical bolometric corrections. Spectrophotometry has been used to simultaneously derive interstellar extinction, effective temperatures and angular diameters for some dEBs in the Large Magellanic Cloud (Guinan \etal\ 1998, Fitzpatrick \etal\ 2002). Knowledge of the radii of the stars then gives the distance to the system, but this method is still being developed  (Groenewegen \& Salaris 2001). A more direct method uses calibrations of photometric indices to derive the surface brightnesses of early-type dEBs (Lacy 1977). Individual Str\"omgren indices for both stars in a dEB can be calculated from the combined system values and the relative brightnesses of the two component stars found in analysis of light curves (Salaris \& Groenewegen 2002). Calibrations exist between the dereddened Str\"omgren index $(b-y)_0$ and visual surface brightness parameter $F_V$ (Barnes \& Evans 1976), derived empirically using interferometric stellar angular diameters (Moon 1984). Accurate calibrations also exist using the broad-band ($V-K$) index (di\,Benedetto 1998). Knowledge of the radii of the components of the dEB gives an entirely empirical distance to each star, potentially accurate to five percent for one binary system. This allows determination of the distance to an open cluster containing a dEB, independent of main sequence fitting, and can be used to calibrate the zero point of other distance indicators, e.g., the three $\delta$ Cephei stars present in NGC\,7790 from the dEB QX\,Cassiopeiae (Sandage 1958).

Probably the best known dEB in an open cluster is V818\,Tauri (HD\,27130, vB\,22) in the Hyades. This was studied by McClure (1982) to derive a Hyades distance modulus in close agreement with the present Hipparcos value. Schiller \& Milone (1987) used V818\,Tau to rederive the Hyades distance modulus and to investigate the mass--luminosity relation of the cluster. Most recently, Pinsonneault \etal\ (2003) compared V818\,Tau to their best-fitting Hyades theoretical isochrone and concluded that the Schiller \& Milone radii of the two stars were inconsistent with the flux ratios. The secondary minimum of this dEB is very shallow and Schiller \& Milone were working with low-quality data, suggesting that the radii are inaccurate and that the quoted errors are internal values from the light curve fitting code. Internal 
errors are known to be optimistic (Popper 1986).

Schiller \& Milone (1988) studied DS\,Andromedae in the nearby old open cluster NGC\,752 to determine the mass and radius at the main sequence turn-off. However, this totally-eclipsing system has an awkward orbital period of 1.01 days and the absolute dimensions of Schiller \& Milone are quite uncertain. 

Other dEBs in open clusters which have been investigated include V906\,Scorpii in NGC\,6475 (M\,7) (Alencar \etal\ 1997) and V392\,Carinae in NGC\,2516 (Debernardi \& North 2001). Many more have been discovered by open cluster variable star searches primarily intended to find other types of variable star, e.g., the surveys of NGC\,6791 by Rucinski, Ka{\l}u\.zny \& Hilditch (1996) and of Collinder\,261 by Mazur, Krzemi\'nski \& Ka{\l}u\.zny (1995).

\subsection{V615\,Persei and V618\,Persei}      \label{discoveryinfo}

Our first paper in this series presents accurate spectroscopic orbits for two well-detached binary systems in h\,Per (NGC\,869), which with $\chi$\,Per (NGC\,884) makes the Perseus Double Cluster. V615\,Per was noted to be variable, possibly of eclipsing nature, by Oosterhoff (1937) but the type of variation of both systems was formally established by Krzesi\'nski, Pigulski \& Ko\l aczkowski (1999, hereafter KPK99). These authors found two primary and two secondary eclipses in the light curve of V615\,Per from over one hundred hours of $UBVI$ observations. They estimated that the period is 13.7136 days. The eclipses are 0.6 and 0.4\,mag deep. KPK99 observed one primary and one secondary eclipse of V618\,Per approximately sixteen days apart, of depths 0.5 and 0.2\,mag, respectively. This did not allow determination of the period, but KPK99 suggested a most likely period of 6.361 days, based on the width of the eclipses and assuming a circular orbit. Our observations give a period of 6.366696 days (see section~\ref{v618period}).

Observed photometric properties for both dEBs are given in Table~\ref{photpartable}. It is notable that the photometric spectral types for both dEBs, B8 and A3, and the orbital periods, 13.7 and 6.4 days, indicate that all four stars are well separated from their companions, so we can be certain that these stars have had completely negligible interaction during their main sequence lifetimes. This is important when comparing individual properties of dEBs to single-star theoretical models. The membership of both systems to the h\,Per open cluster is also in little doubt. They fit onto the
binary main sequence in all cluster colour--magnitude diagrams (CMDs), are situated on the sky in the cluster nucleus, and V615\,Per has a measured proper motion which implies a membership probability of 0.96 (Uribe \etal\ 2002). Both dEBs
were considered to be members of h\,Per by van\,Maanen (1944) from a study of the proper motions of the stars in the region of h and $\chi$ Persei.

\subsection{h\,Persei and $\chi$\,Persei}     \label{clusterinfo}

\begin{table*} \begin{center} \caption{\label{clusterpartable} Selected values of distance modulus, age and reddening taken from the literature. If one value is quoted for both clusters it is included in the table between the two relevant columns. Findings of differential reddening have not, in general, been indicated, and in such cases a best single reddening has been quoted. \newline $^*$\,Reddening values in the Str\"omgren system have been converted to broad-band indices using $\EBV \approx 1.37 \Eby$ (Crawford 1975). \newline $^\dag$\,Converted from the Geneva photometric reddening index $E[B-V]$ by Waelkens \etal\ (1990) using $\EBV \approx 0.86 E[B-V]$.}
\begin{tabular}{|l|cc|cc|cc|} \hline \hline
Reference                     & \multicolumn{2}{|c|}{Distance modulus} & \multicolumn{2}{|c|}{$\log \tau$ (years)} 
\                             & \multicolumn{2}{|c|}{Reddening \EBV} \\
\                             & h\,Per & $\chi$\,Per & h\,Per & $\chi$\,Per & h\,Per & $\chi$\,Per \\ \hline
Oosterhoff (1937)             & \multicolumn{2}{|c|}{11.51} & & & & \\
Bidelmann (1943)              & \multicolumn{2}{|c|}{11.42} & & & & \\
Johnson (1957)                & \multicolumn{2}{|c|}{11.76} & & & & \\
Wildey (1964)                 & \multicolumn{2}{|c|}{11.9 } & \multicolumn{2}{|c|}{$6.78 - 7.78$} & \\
Schild (1967)                 & 11.66 & 11.99 & 6.81 & 7.06 & \\
Crawford \etal\ (1970)        & \multicolumn{2}{|c|}{$11.4 \pm 0.4$} &  &  & \multicolumn{2}{|c|}{$0.56 \pm 0.03^*$} \\
Balona \& Shobbrook (1984)    & \multicolumn{2}{|c|}{$11.17 \pm 0.09$} & & & & \\
Tapia \etal\ (1984)           & & & & & $0.58 \pm 0.03$ & $0.59 \pm 0.03$ \\
Liu \etal\ (1989)             & 11.74 & 11.73 & 7.26 & 6.48 & & \\
Waelkens \etal\ (1990)        & & & & & $0.56 \pm 0.03^\dag$ & $0.56 \pm 0.06$\dag \\     
Krzesi\'nski \etal\ (1999)    & & & & & \multicolumn{2}{|c|}{0.52} \\
Marco \& Bernabeu (2001)      & $11.56 \pm 0.20$ & $11.66 \pm 0.20$ & $6.8 - 7.0$ & 7.15, 7.3 
\                                         & $0.44 \pm 0.02^*$ & $0.39 \pm 0.05^*$\\
Keller (2001)                 & \multicolumn{2}{|c|}{$11.75 \pm 0.05$} & \multicolumn{2}{|c|}{$7.10 \pm 0.01$} 
\                             & \multicolumn{2}{|c|}{$0.54 \pm 0.02$} \\
Uribe (2002)                  & $11.42 \pm 0.09$ & $11.61 \pm 0.06$ & & & & \\    
Slesnick (2002)               & \multicolumn{2}{|c|}{$11.85 \pm 0.05$} & $7.10 \pm 0.01$ & $7.11 \pm 0.01$ 
\                                         & $0.57 \pm 0.08$ & $0.53 \pm 0.08$\\
Capilla \& Fabregat(2002)     & \multicolumn{2}{|c|}{$11.7 \pm 0.1$} & \multicolumn{2}{|c|}{$7.10 \pm 0.05$}
\                                         & 0.449--0.637$^*$ & $0.545 \pm 0.034^*$ \\
\hline \hline \end{tabular} \end{center} \end{table*}

The Perseus Double Cluster is a rich, young open cluster system relatively close to the Sun. This has made it an important and frequently used tool for studying the evolution of massive stars, and it is one of the most studied objects in the Northern Hemisphere. Many studies have been motivated by the disputed connection between h\,Per and $\chi$\,Per. Their proximity to each other and the similar morphology of their photometric diagrams has led to the suggestion that the clusters are co-evolutionary, and in this sense perhaps unique in the Milky Way (see also Sandage 1958). The Double Cluster is also traditionally taken to be nucleus of the Perseus\,OB1 association (Humphreys 1978) although Slesnick \etal\ (2002) argue that it is impossible to be certain using current observational techniques. 

The first detailed study was undertaken by Oosterhoff (1937), who used photographic photometry and very low-resolution photographic spectrophotometry to assign an ``effective wavelength'' to each star studied. Wildey (1964) conducted extensive photoelectric photometry of the general area and ascribed ages of 7, 17 and 60 Myr to the turn-off morphology of three perceived main sequences in the cluster CMD. He also found ages of 6\,Myr for pre-main-sequence stars and at least 46\,Myr for the faintest main sequence star observed. Schild (1965, 1967) claimed differences between the CMD morphology of the two clusters in the sense that h\,Per was older than $\chi$\,Per and 0.3\,mag more distant, even allowing for a 0.2\,mag difference in extinction. He also noted that $\chi$\,Per contained many Be stars whilst h\,Per did not, implying significant evolutionary differences. 

Crawford, Glaspey \& Perry (1970) observed the clusters in the Str\"omgren system and claimed there was no evidence that the clusters were not co-evolutionary. Waelkens \etal\ (1990) observed the cluster nuclei in the intermediate-band Geneva photometric system and confirmed the conclusions of Crawford \etal\ Intermediate-band systems have better procedures for individually dereddening single stars. This capability is important for clusters, such as h\,Per, which display differential reddening. This may be the reason why intermediate-band photometric studies (before the year 2000) tend to find that h and $\chi$ Per have common properties whereas broad-band studies do not.

Tapia \etal\ (1984) conducted $JHK$ photometry and suggested that the variable reddening found in many previous studies may not be interstellar but intrinsic to the atmospheres of some B\,stars. They found no variation in extinction over the cluster but stated that a significant difference exists in the stellar contents of the two clusters, casting doubt on their co-evolutionary status.

There have been four recent photometric studies of the Double Cluster. Str\"omgren data were taken by Marco \& Bernabeu (2001) who claimed that there were three distinct epochs of star formation: one of 6.3 to 10 Myr in h\,Per, and two of 14 and 20 Myr in $\chi$\,Per. The distance moduli derived were consistent with a common distance.

Broadband observations were published by Keller \etal\ (2001) and Slesnick \etal\ (2002). Both studies found a common distance and age for h and $\chi$ Per. Keller \etal\ claimed that Marco \& Bernabeu had overinterpreted their data whilst Slesnick \etal\ claimed that Wildey (1964) did not sufficiently consider contamination by field stars, particularly background late-type giants.

Capilla \& Fabregat (2002) undertook more extensive Str\"omgren photometry than Marco \& Bernabeu and claim a common distance and age for h\,Per and $\chi$\,Per. They also, like many previous studies, find strong differential reddening over h\,Per and weaker, constant reddening over $\chi$\,Per. Comparison of their observed and de-reddened photometric diagrams strongly implies that differences in reddening and membership selection have been the main cause of dispute over the relative and absolute physical status of the two open clusters. 

Table~\ref{clusterpartable} lists selected published parameters of the two clusters. If the last four photometric studies are considered it can be seen that the values are converging towards a distance modulus of $11.70 \pm 0.05$ and an age of $\log\tau = 7.10 \pm 0.01$ (years). We will adopt these values for purposes of discussion and model comparison
in this paper.


\section{Observations}  \label{observations}  

\subsection{Spectroscopy}

Spectroscopic observations were carried out in 2002 October using the 2.5\,m Isaac Newton Telescope (INT) on La Palma. The 500\,mm camera of the Intermediate Dispersion Spectrograph (IDS) was used with a holographic 2400\,{\it l}\,mm$^{-1}$ grating. An EEV 4k\,$\times$\,2k CCD was used and exposure times were 1800 seconds. From measurements of the full width half maximum (FWHM) of arc lines taken for wavelength calibration we estimate that the resolution is approximately 0.2\,\AA. The main spectral window chosen for observation was 4230--4500\,\AA. This contains the \ion{Mg}{II}\ 4481\,\AA\ line which is known to be one of the best lines for radial velocity work for early-type stars (Andersen 1975; Kilian, Montenbruck \& Nissen 1991). \ion{He}{I}\ 4471\,\AA\ and H$\gamma$ (4340\,\AA) are useful for determination of effective temperatures and spectral types for such stars. One spectrum of V615\,Per was observed at H$\beta$ (4861\,\AA) to provide an additional temperature indicator. The spectra of V615\,Per have an average signal to noise per pixel (S/N) of approximately 50, whereas the S/N for the spectra of V618\,Per is approximately 15. 

Data reduction was undertaken using optimal extraction as implemented in the software tools {\sc pamela} and {\sc molly} \footnote{{\sc pamela} and {\sc molly} were written by Dr.\ Tom Marsh and are found at \texttt{http://www.astro.soton.ac.uk/$^{\sim}\!$trm/software.html}} (Marsh 1989).

\subsection{Photometry}

Observations in the intermediate-band Str\"omgren $uvby$ and Crawford $\beta$ system were undertaken at the 1\,m Jakobus
Kapteyn Telescope (JKT), also on La Palma, during 2002 December and 2003 January. The $uvby\beta$ photometric system was
designed to provide accurate photometric parameters for early-type stars (Str\"omgren 1966) and is useful in this case for its robust procedures concerning interstellar reddening (see Crawford 1978, 1979), which is known to be large and variable towards the Perseus Double Cluster.

Light curves were observed in the $b$ and $y$ filters and are complete over the primary and secondary eclipses of both systems. Reduction was undertaken using aperture photometry due to slight charge transfer problems. This caused photometry based on analysis of the point spread function to be unreliable. Light curves from some nights exhibit significant night errors, and whilst the internal precision of the data is good, observations on different nights fail to agree on the outside-eclipse brightness of the system by about 0.05\,mag. This is not due to intrinsic variability: such an effect is not present in the discovery light curves from KPK99 but was noticed in other data obtained on the same observing run as V615\,Per and V618\,Per.

A nonlinearity was also found in the CCD images. This was quantified by fitting a polynomial to the magnitudes of the stars on each image compared to the magnitudes of the same stars on a reference image. Removal of the nonlinearity effects halved the night errors but the light curves of V615\,Per are unsuitable for model fitting. The data for V618\,Per seem to be less affected, and the effect can be minimised by offsetting light curves from different nights by small amounts (of the order of 0.01\,mag).

\begin{table} \begin{center} \caption{\label{v615mintable} 
Times of minima and $O-C$ values determined for V615\,Per from data taken with the JKT. Cycle zero was chosen to be the eclipse with the best-defined time of minimum. The times of minimum for cycle 36.0 are incorrect by approximately five minutes and were not used to determine the period. They are included here for completeness.
\newline $^\dag$\,All times are given as (${\rm HJD} - 2\,400\,000$). 
\newline $^\ddag$\,The quoted error is the formal error of the Gaussian fit.}
\begin{tabular}{l l r r@{.}l c r} \hline \hline
Source & & Cycle & \multicolumn{2}{c}{$T_0$ (HJD) $^\dag$} & Error $^\ddag$ & $O-C$ \ \\ \hline
2001 Sep  & $B$    &   0   &  52169&68218    & 0.00085  &    0.00008   \\
2001 Sep  & $V$    &   0   &  52169&68200    & 0.00088  & $-$0.00010   \\
2001 Sep  & $I$    &   0   &  52169&68218    & 0.00078  &    0.00008   \\
2002 Dec  & $b$    &  32.5 &  52615&38403    & 0.00034  &    0.00018   \\
2002 Dec  & $y$    &  32.5 &  52615&38316    & 0.00031  & $-$0.00069   \\
2003 Jan  & $b$    &  35.5 &  52656&52649    & 0.00054  &    0.00094   \\
2003 Jan  & $y$    &  35.5 &  52656&52509    & 0.00050  & $-$0.00046   \\
2003 Jan  & $b$    &  36   &  52663&37978    & 0.00032  & $-$0.00272   \\
2003 Jan  & $y$    &  36   &  52663&37950    & 0.00027  & $-$0.00300   \\
\hline \hline \end{tabular} \end{center} \end{table}

The KPK99 discovery light curves contain a total coverage of 24, 57, 103 and 10 hours of observation in the broad-band $U$, $B$, $V$ and $I$ filter respectively. Although observations are somewhat sparse during eclipses of V615\,Per and V618\,Per, the light curves are of sufficient quality for an approximate determination of the stellar radii and orbital inclination of V615\,Per. They also show no sign of any stellar brightness variation apart from the eclipses.

Supplementary $BVI$ service data were taken with the JKT on 2001 September 16 to capture part of a primary eclipse of V615\,Per. This also serendipitously captured a descending branch of a primary eclipse of V618\,Per. These data were reduced using optimal extraction as implemented in the Starlink package {\sc autophotom}\footnote{The Starlink software world wide web homepage is \texttt{http://www.starlink.rl.ac.uk/}} and combined with the KPK99 discovery light curves. We consider this slight inhomogeneity of the $BVI$ data to be negligible, and comparable to the inhomogeneity introduced by KPK99 by the use of two different observatories in their search for variable stars.


\section{Period determination} \label{perioddet}      

\subsection{V615 Per} \label{v615period}

KPK99 observed descending and ascending branches of two primary and two secondary eclipses and determined a period for 
V615\,Per of 13.7136(1) days, where the number in parentheses refers to uncertainty in the last digit of the value 
given. As no observations overlapped during eclipse, and no actual light minima were observed, it was not possible to 
determine the period of light variation without assuming a certain shape for both primary and secondary eclipses. This 
difficulty resulted in them underestimating the width of the eclipses and their ephemeris disagrees slightly with our 
own.

The times of mid-eclipse are reproduced in Table~\ref{v615mintable} for those eclipses for which an actual light minimum was observed. The $BVI$ light curves obtained as service data were fitted with Gaussian functions to determine times of minimum and their formal errors. Gaussian functions provide a very good representation of the eclipse shapes of this system as the eclipses are deep but not total, and the orbit has a very small eccentricity so eclipses will be symmetric about their centre. The night errors in the Str\"omgren $b$ and $y$ light curves may cause asymmetry and so bias the results derived by fitting a Gaussian function, so only the central parts of the eclipse were fitted. 

A straight line fitted by least squares to the times of minima, using the most accurately determined time of eclipse as cycle zero, showed larger $O-C$ (observed minus calculated) values than expected. Inspection of phased light curves indicated that the correct period had to be 13.71390\,d so the times of minimum for cycle 36.0 are earlier than expected. A straight line fit to the remaining times of minima gives the ephemeris \[ {\rm Min\,I} = {\rm HJD}\,2\,452\,169.6821(5) + 13.71390(2)\,\times\,E \] The above ephemeris will be used throughout this work.

\subsection{V618 Per}   \label{v618period}

\begin{table} \begin{center} \caption{\label{v618mintable}
Times of minima and $O-C$ values determed for V618\,Per. Cycle zero was chosen to be the eclipse with the best-defined time of minimum. Only times of primary minima were used to determine the final ephemeris.
\newline $^\dag$\,All times are given as (${\rm HJD} - 2\,400\,000$). 
\newline $^\ddag$\,The quoted error is the formal error of the Gaussian fit.}
\begin{tabular}{l l r r@{.}l c r} \hline \hline
Source & & Cycle & \multicolumn{2}{c}{$T_0$ (HJD) $^\dag$} & Error $^\ddag$ & $O-C$ \ \\ \hline
KPK99     & $B$    & $-$398   &  50081&4483  &  0.0069  & $-$0.0022  \\
KPK99     & $V$    & $-$398   &  50081&4523  &  0.0021  &    0.0019  \\
KPK99     & $B$    & $-$395.5 &  50097&3528  &  0.0057  & $-$0.0144  \\
KPK99     & $V$    & $-$395.5 &  50097&3541  &  0.0041  & $-$0.0132  \\
2001 Sep  & $B$    &  $-$70   &  52169&7212  &  0.0015  & $-$0.0056  \\
2001 Sep  & $V$    &  $-$70   &  52169&7260  &  0.0016  & $-$0.0008  \\
2001 Sep  & $I$    &  $-$70   &  52169&7232  &  0.0027  & $-$0.0036  \\
2002 Dec  & $b$    &      0   &  52615&3953  &  0.0007  & $-$0.0003  \\
2002 Dec  & $y$    &      0   &  52615&3958  &  0.0010  &    0.0003  \\
2003 Jan  & $b$    &      5.5 &  52650&4161  &  0.0019  &    0.0039  \\
2003 Jan  & $y$    &      5.5 &  52650&4135  &  0.0013  &    0.0012  \\
\hline \hline \end{tabular} \end{center} \end{table}

KPK99 observed one primary and one secondary eclipse of V618\,Per, separated by approximately 16 days. The light curves each cover just over half of one minimum but are very sparse. V618\,Per was also in eclipse during the service observations of the eclipse of V615\,Per and just over half of one primary eclipse was observed in $BVI$. Our $b$ and $y$ light curves covered most of one primary and most of one secondary eclipse but the small depth of the secondary eclipse means that the primary eclipse has a better-defined minimum. 

All eclipses were fitted with Gaussian functions. Eclipse widths were held fixed to the width of the best observed eclipse and the uncertainty generated by this has been added in quadrature with the formal errors of the Gaussian fit. The times of mid-eclipse are reproduced in Table~\ref{v618mintable}. Adopting a period found using a linear least-squares fit to the primary minima and a timebase corresponding to the best-defined light minimum observed (in two filters) gives the ephemeris \[ {\rm Min\,I} = {\rm HJD}\,2\,452\,615.3955(3) + 6.366696(4)\,\times\,E \] This period will be used throughout this paper. The secondary minima contain fewer datapoints and are shallower than the primary minima, but give a period similar to that derived from the primary eclipses.


\section{Spectral disentangling}   \label{disentangling}

\begin{figure} \includegraphics[width=0.48\textwidth,angle=0]{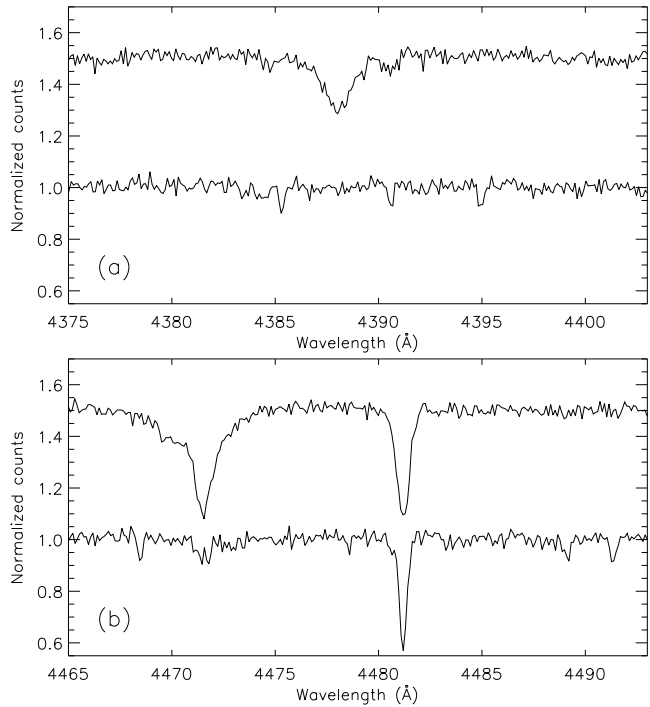} \\
\caption{\label{v615disentangledplot} Disentangled spectra for V615\,Per. Two spectral windows are shown, with the primary spectrum offset by +0.5 for clarity. Panel (a) contains the \ion{He}{I} 4388\,\AA\ line in the primary spectrum and several sharp weak secondary lines. Panel (b) contains the \ion{He}{I} 4471\,\AA\ and the \ion{Mg}{II}\ 4481\,\AA\ lines from which most radial velocity and effective temperature information were derived.} \end{figure}

Spectral disentangling is a method to determine the two individual spectra which best fit a set of composite binary spectra well distributed in orbital phase (Simon \& Sturm 1994). The light ratio of the component stars must be the same
in each observed spectrum (so spectra taken during an eclipse must be rejected) and both stars must show no spectral variability. The best-fitting disentangled spectra can then be found by singular value decomposition (Simon \& Sturm 1994) or by Fourier analysis (Hadrava 1995). Disentangling can be used to determine accurate orbital semiamplitudes (Hynes \& Maxted 1998, Harries \etal\ 2003) but it is not clear if there is a robust method by which to estimate uncertainties in the derived quantities. 

Disentangling derives individual spectra of the components of binary stars, which may then be analysed using spectral synthesis to find the effective temperatures and rotational velocities of the two stars. This information can then be used to generate synthetic spectra needed to determine a spectroscopic orbit with {\sc todcor} (section~\ref{todcor}). As disentangling requires a spectroscopic orbit to calculate the radial velocities for the stars in each observed spectrum, we derived a preliminary orbit for V615\,Per. Gaussian functions were fitted to the \ion{Mg}{II} 4481\,\AA\ spectral lines and a spectroscopic orbit was fitted to the resulting radial velocities using {\sc sbop}\footnote{Spectroscopic Binary Orbit Program written by Dr.\ Paul B.\ Etzel (\texttt{http://mintaka.sdsu.edu/faculty/etzel/}).}. The results are consistent with a circular orbit (the eccentricity value found is smaller than its standard error) so a final solution was made with no eccentricity. The velocity semiamplitudes are $K_{\rm A} = 75.9 \pm 0.8$\kms\ and $K_{\rm B} = 95.9 \pm 0.7$\kms.

The Simon \& Sturm algorithm was used to produce separate spectra of the components of V615\,Per. Such spectra are sufficiently reliable for quantitative spectral analysis using line strength ratios, although absolute line strengths may be less reliable as disentangling is independent of the binary light ratio. The resulting spectra show significant variation in continuum level over the observed wavelength range. This is easily removed by polynomial fitting over small
spectral windows but cannot cope with the shapes of broad lines, so the disentangled spectra have unreliable H$\gamma$ 4340\,\AA\ profiles. The individual spectra are shown in two spectral windows in Fig.~\ref{v615disentangledplot}.

A preliminary spectroscopic orbit was found for V618\,Per by disentangling the observed spectra over a grid of values of $K_{\rm A}$ and $K_{\rm B}$ to find the values for which the residuals of the fit were the lowest.


\section{Spectral synthesis} \label{specsynth}  

\begin{figure} \includegraphics[width=0.48\textwidth,angle=0]{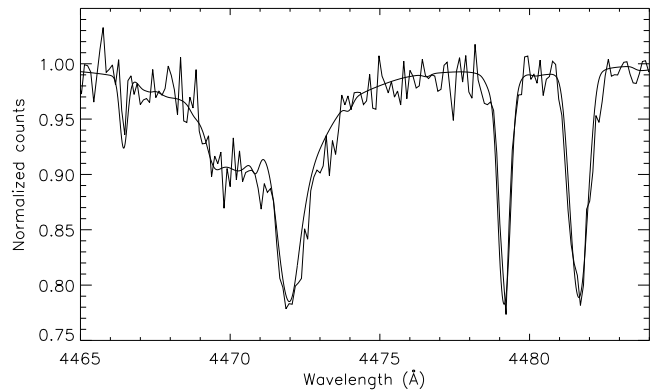} \\ \caption{\label{v615specsynth} 
Representation of the best-fitting synthetic composite spectrum of V615\,Per. The thick line shows the average of the last four spectra observed on HJD\,2\,452\,564. This has less noise than one spectrum and the orbital smearing of the spectral lines is approximately 3\kms\ for both stars. The spectral window containing the \ion{He}{I} 4471\,\AA\ and \ion{Mg}{II}\ 4481\,\AA\ lines is shown and the primary lines are redward of the secondary lines.} \end{figure}

Effective temperatures and rotational velocities were derived for V615\,Per by comparing the observed and disentangled spectra with synthetic spectra calculated using {\sc uclsyn} (Smith 1992, Smalley \etal\ 2001), Kurucz (1993) {\sc atlas9} model atmospheres and absorption lines from the Kurucz \& Bell (1995) linelist. The profiles for the 4387.93\,\AA\ and 4471.50\,\AA\ \ion{He}{I}\ lines were calculated using profiles from the work of Barnard \etal\ (1969)
and Shamey (1969), with $\log gf$ values from the critical compilation of Wiese \etal\ (1966). The spectra were rotationally broadened as necessary and instrumental broadening was applied with FWHM\,=\,0.2\,\AA\ to match the resolution of the observations. 

V615\,Per was spectroscopically analysed using the binary star mode ({\sc binsyn}) within {\sc uclsyn}. A value of $\log g = 4.4$ was adopted for both components, based on preliminary analyses. Microturbulence velocities of 0\kms\ and 2\kms\ were assumed for the primary and secondary, respectively. Properties of the two components were obtained by fitting to the observations using the least-square differences, which also enabled a monochromatic light ratio to be obtained. 

The effective temperature and rotational velocity derived for the primary are $15000 \pm 500$\,K and $v \sin i = 28 \pm 5$\kms\ respectively. For the secondary these values are $11000 \pm 500$\,K and $8 \pm 5$\kms\ respectively. The relative contributions of the stars to the total system light at a wavelength of 4481\,\AA\ are $0.65 \pm 0.03$ and $0.35 \mp 0.03$ for the primary and secondary respectively. These results are robust against small changes in metallicity but rely on the helium abundance being roughly solar.

\begin{table} \begin{center} \caption{\label{v615rvtable}
Radial velocities and $O-C$ values (in \kms) for V615\,Per calculated using {\sc todcor}. Weights were derived from the amount of light collected in that observation and were used in the {\sc sbop} analysis.}
\begin{tabular}{lrrrrr} \hline \hline
HJD $-$      & Primary    & $O-C$   & Secondary & $O-C$   & Wt   \\ 
2\,400\,000  & velocity   &         & velocity  &         &      \\ \hline
52561.7284   &    $-$3.4  & $-$1.3 \ &  $-$91.6  &    2.2 \ & 1.5  \\
52562.4520   &      17.6  &    1.6 \ & $-$118.6  & $-$0.5 \ & 1.2  \\
52562.5235   &      20.1  &    2.7 \ & $-$118.6  &    1.5 \ & 1.0  \\
52562.6300   &      22.5  &    3.0 \ & $-$122.9  &    0.0 \ & 0.9  \\
52563.4559   &      28.5  & $-$1.4 \ & $-$142.0  & $-$4.0 \ & 0.9  \\
52563.5074   &      27.9  & $-$2.3 \ & $-$136.5  &    2.0 \ & 0.9  \\
52563.5754   &      31.7  &    1.2 \ & $-$143.8  & $-$4.6 \ & 0.5  \\
52563.6576   &      32.3  &    1.5 \ & $-$141.4  & $-$1.7 \ & 0.2  \\
52564.4086   &      21.0  & $-$7.8 \ & $-$137.1  &    1.8 \ & 1.1  \\
52564.5128   &      27.6  & $-$0.2 \ & $-$138.9  & $-$1.0 \ & 1.2  \\
52564.5602   &      30.0  &    2.6 \ & $-$137.4  & $-$0.0 \ & 1.1  \\
52564.6517   &      28.8  &    2.5 \ & $-$136.2  &    0.0 \ & 1.2  \\
52564.6958   &      30.3  &    4.5 \ & $-$139.2  & $-$3.6 \ & 0.5  \\
52565.3995   &       8.9  & $-$4.4 \ & $-$120.3  &    0.7 \ & 1.1  \\
52565.4713   &      12.0  &    0.4 \ & $-$118.9  &    0.1 \ & 1.1  \\
52565.5478   &      11.4  &    1.7 \ & $-$117.4  & $-$0.6 \ & 0.8  \\
52565.5959   &       9.7  &    1.1 \ & $-$115.4  &    0.0 \ & 0.6  \\
52566.3911   &   $-$16.7  & $-$2.6 \ &  $-$78.9  &    8.6 \ & 0.3  \\
52568.6758   &   $-$82.4  &    6.6 \ &      6.5  & $-$2.5 \ & 0.6  \\
52569.5196   &  $-$109.6  & $-$1.0 \ &     36.3  &    0.7 \ & 1.0  \\
52569.6165   &  $-$115.7  & $-$5.4 \ &     35.5  & $-$2.5 \ & 0.1  \\
52570.4977   &  $-$120.0  & $-$0.7 \ &     51.4  & $-$0.1 \ & 1.7  \\
52570.5873   &  $-$118.0  &    1.5 \ &     54.3  &    2.2 \ & 1.3  \\
52570.6997   &  $-$119.2  &    0.5 \ &     52.3  & $-$0.2 \ & 1.2  \\
52571.5213   &  $-$122.6  & $-$8.0 \ &     39.3  & $-$8.6 \ & 0.4  \\
\hline \hline \end{tabular} \end{center} \end{table}

\begin{table} \begin{center} \caption{\label{v618rvtable}  
Radial velocities and $O-C$ values (in \kms) for V615\,Per calculated using {\sc todcor}. Weights were derived from the amount of light collected in that observation and were used in the {\sc sbop} analysis.
\newline $^\dag$\,Radial velocities rejected from {\sc sbop} fit (see text for details).}
\begin{tabular}{lr@{}lrr@{}lrr} \hline \hline
HJD $-$     & \multicolumn{2}{r}{Primary}  & $O-C$ & \multicolumn{2}{r}{Second.}  & $O-C$ & Wt \\ 
2\,400\,000 & \multicolumn{2}{r}{velocity} &       & \multicolumn{2}{r}{velocity} &       &    \\ \hline
52559.4522  & $-$113.4 &        &     1.4 \ &     63.9 &         &     2.9 \ & 1.4    \\
52559.4733  & $-$115.6 &        &  $-$0.5 \ &     66.9 &         &     5.4 \ & 1.3    \\
52559.6311  & $-$117.6 &        &  $-$1.0 \ &     63.9 &         &     0.2 \ & 1.0    \\
52559.6522  & $-$118.1 &        &  $-$1.5 \ &     62.0 &         &  $-$1.8 \ & 1.0    \\
52560.4354  & $-$104.8 &        &  $-$6.9 \ &     28.2 &         &  $-$7.5 \ & 0.7    \\
52560.5872  &  $-$93.6 &        &  $-$3.6 \ &     25.5 &         &     1.5 \ & 0.4    \\
52561.5565  &  $-$22.5 &        &     2.4 \ &  $-$78.6 &         &  $-$5.0 \ & 1.5    \\
52561.6385  &  $-$22.7 &        &  $-$3.5 \ &  $-$81.2 &         &     0.7 \ & 1.8    \\
52562.4759  &     23.6 &        &     1.2 \ & $-$144.7 &         &  $-$0.3 \ & 1.3    \\
52562.5471  &     24.8 &        &     0.6 \ & $-$147.6 &         &  $-$0.6 \ & 1.2    \\
52562.6536  &     22.2 &        &  $-$4.1 \ & $-$148.4 &         &     1.6 \ & 1.5    \\
52562.7104  &     28.2 &        &     1.3 \ & $-$152.8 &         &  $-$1.7 \ & 1.5    \\
52563.4795  &     16.1 &        &     0.9 \ & $-$133.1 &         &     0.4 \ & 1.1    \\
52563.5485  &     14.0 &        &     1.8 \ & $-$129.0 &         &     0.1 \ & 0.8    \\
52563.5993  &     10.3 &        &     0.4 \ & $-$123.5 &         &     2.1 \ & 1.0    \\
52563.6796  &      6.2 &        &     0.2 \ & $-$120.0 &         &  $-$0.3 \ & 0.9    \\
52563.7344  &      4.2 &        &     1.0 \ & $-$116.3 &         &  $-$0.8 \ & 0.8    \\
52564.4324  &  $-$43.0 &        &  $-$0.7 \ &  $-$48.8 &         &  $-$1.3 \ & 1.0    \\
52564.5839  &  $-$50.8 &        &     2.3 \ &  $-$35.8 &         &  $-$4.5 \ & 1.6    \\
52564.6759  &  $-$53.8 &        &     5.8 \ &  $-$18.4 &         &     3.3 \ & 1.4    \\
52565.4231  & $-$103.6 &        &  $-$0.4 \ &     43.9 &         &     0.3 \ & 1.6    \\
52565.4942  & $-$103.0 &        &     3.0 \ &     48.5 &         &     0.7 \ & 1.2    \\
52565.5745  & $-$109.9 &        &  $-$1.1 \ &     49.1 &         &  $-$2.9 \ & 0.5    \\
52565.6207  & $-$109.3 &        &     0.9 \ &     52.6 &         &  $-$1.6 \ & 0.7    \\
52566.4145  & $-$159.1 &$^\dag$ & $-$47.0 \ &     56.6 &         &  $-$0.5 \ & 0.4    \\
52568.7038  &     30.0 &$^\dag$ &    11.9 \ & $-$133.1 &         &     4.7 \ & 0.2    \\
52569.4302  &     23.6 &        &  $-$2.9 \ & $-$149.6 &         &     0.9 \ & 1.1    \\
52569.5426  &     66.2 &$^\dag$ &    41.6 \ & $-$145.5 &         &     2.1 \ & 1.0    \\
52569.6761  &     21.6 &        &     0.4 \ & $-$140.0 &         &     2.4 \ & 0.9    \\
52571.6157  &  $-$98.1 &        &  $-$3.1 \ &  $-$29.7 &$^\dag$  & $-$61.3 \ & 1.0    \\
\hline \hline \end{tabular} \end{center} \end{table}

The spectra of V618\,Per are of much lower S/N so a wide range of parameters provided acceptable fits. Microturbulence velocities of 0\kms\ and surface gravities of $\log g = 4.4$ were assumed for both components. The effective temperatures and rotational velocities found are $11000 \pm 1000$\,K and $10 \pm 5$\kms\ for the primary and $8000 \pm 1000$\,K and $10 \pm 5$\kms\ for the secondary. The relative contributions of the stars to the total system light are $0.7 \pm 0.1$ and $0.3 \mp 0.1$.


\section{Spectroscopic orbits} \label{todcor}   

\begin{table*} \begin{center} \caption{\label{v615v618orbitdata} Final spectroscopic orbit for both dEBs using {\sc sbop} to fit radial velocities derived from {\sc todcor}. All symbols have their usual meanings and those parameters held fixed in the {\sc sbop} analysis are indicated. All quoted uncertainties include errors arising from spectral template mismatch, added in quadrature. The ephemeris timebase $T_0$ refers to the time of minimum light of a primary eclipse.}
\begin{tabular}{l r@{\,$\pm$\,}l c r@{\,$\pm$\,}l c r@{\,$\pm$\,}l c r@{\,$\pm$\,}l} \hline \hline
\hspace*{70pt}       & \multicolumn{2}{c}{V615\,Per A} & \hspace*{10pt} & \multicolumn{2}{c}{V615\,Per B} & 
\hspace*{20pt} & \multicolumn{2}{c}{V615\,Per A} & \hspace*{10pt} & \multicolumn{2}{c}{V618\,Per B} \\ \hline
Period (days)        & \multicolumn{5}{c}{13.71390 \ (fixed)}     &  & \multicolumn{5}{c}{6.366696 \ (fixed)} \\
Ephemeris $T_0$ (HJD)&\multicolumn{5}{c}{2\,452\,169.6821 \ (fixed)}&&\multicolumn{5}{c}{2\,452\,615.3955 \ (fixed)}\\
Velocity semiamplitude $K$ (\kms)  & 75.44 & 0.82 & & 96.71 & 0.62 & & 72.28 & 0.82 & & 108.16 & 0.69 \\
Systemic velocity (\kms) & $-$44.27 & 0.73 & & $-$44.08 & 0.54 & & $-$44.42 & 0.82 & & $-$44.29 & 0.51 \\
Orbital eccentricity & \multicolumn{5}{c}{0.0 \ (fixed)}          &  & \multicolumn{5}{c}{0.0 \ (fixed)} \\ \hline
$M \sin^3 i$ (\Msun) & 4.072     & 0.055 &   & 3.177     & 0.051  &  & 2.323     & 0.031  &  & 1.552     & 0.025 \\
$a \sin i$ (\Rsun)   & \multicolumn{5}{c}{46.64 $\pm$ 0.28}       &  & \multicolumn{5}{c}{22.70 $\pm$ 0.13} \\
Mass ratio $q$       & \multicolumn{5}{c}{0.7801 $\pm$ 0.0098}    &  & \multicolumn{5}{c}{0.6682 $\pm$ 0.0087} \\
\hline \hline \end{tabular} \end{center} \end{table*}

Cross-correlation of observed spectra against a template spectrum (either synthetic or from a standard star) is a powerful technique for the determination of radial velocities. In composite spectra, however, the calculated position of greatest correlation for one star can be shifted by the presence of the second star in the spectrum. The two-dimensional cross-correlation algorithm {\sc todcor} (Zucker \& Mazeh 1994) compensates for this by fitting templates for both stars simultaneously. This reduces the effects of spectral line blending and also allows different templates to be used for primary and secondary stars. The choice of input templates to correlate observed spectra against is important. The best results are obtained for spectra which most closely match those of the stars under investigation, and differences in effective temperature, surface gravity and rotational velocity can cause random and systematic errors in the calculated radial velocities. Using spectra of standard stars avoids the use of models but increases random error (as the template spectra contain observational noise) and limits the choice of spectrum. Synthetic spectra are generally preferred as they are easily obtained for any combination of effective temperature, surface gravity and rotational velocity. They are also free of noise but can introduce a bias into the results due to their differences from observed spectra, e.g., missing spectral lines.

\subsection{V615 Per}

\begin{figure} \includegraphics[width=0.48\textwidth,angle=0]{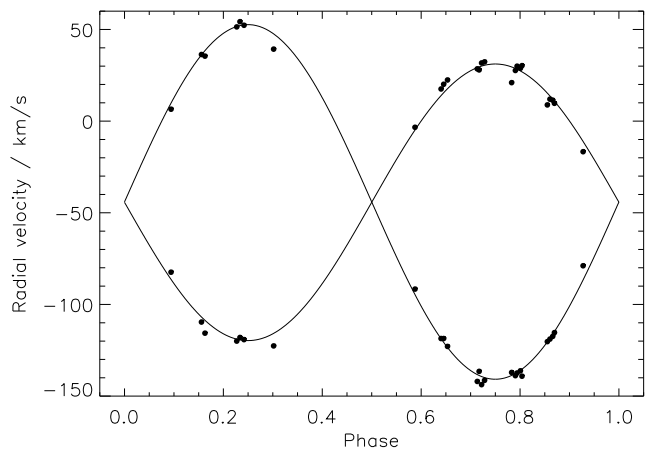} \\ \caption{\label{v615orbitplot} 
Spectroscopic orbit for V615\,Per from an {\sc sbop} fit to radial velocities from {\sc todcor}.} \end{figure}

Radial velocities were obtained using the {\sc todcor} algorithm. The template spectra used were synthetic spectra generated with the effective temperatures and rotational velocities found in the spectral synthesis analysis (section~\ref{specsynth}). For V615\,Per the effects of template mismatch on the derived velocity semiamplitudes are well below the standard errors of the fit and have been found to be 0.1, 0.05, 0.05 and 0.01\kms, respectively for effective temperature, rotational velocity, microturbulence velocity and the size of a mask positioned over the broad H$\gamma$ 4340\,\AA\ line. These have been added in quadrature to the final errors on the velocity semiamplitudes derived. 

\begin{figure} \includegraphics[width=0.48\textwidth,angle=0]{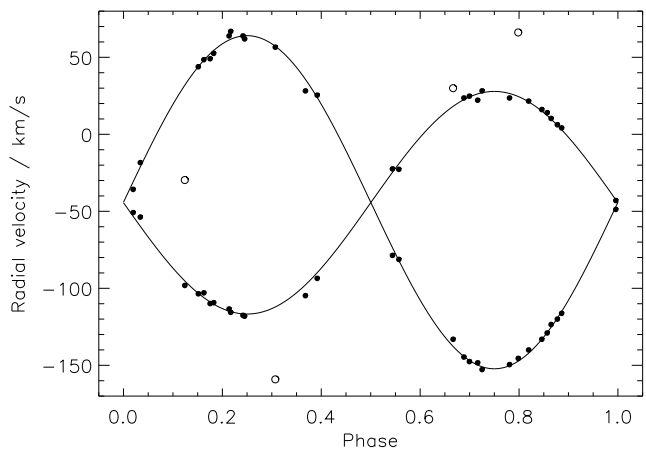} \\
\caption{\label{v618orbitplot} Spectroscopic orbit for V618\,Per from the {\sc todcor} analysis. Filled circles indicate radial velocities included in the {\sc sbop} fit and open circles indicate rejected radial velocities} \end{figure}

The radial velocities have been reproduced in Table~\ref{v615rvtable} and fitted with {\sc sbop} in single-lined Lehmann Filh\'es mode. It is known that the two components of a spectroscopic binary can have slightly inconsistent systemic velocities due to physical effects (such as gas streams affecting spectroscopic lines or the use of different spectral features for the two stars) or from neglect of a small eccentricity (Popper 1974). Analysis of the light curves of V615\,Per suggest a possible eccentricity (see section~\ref{v615ebop}) too small to detect in our spectroscopy. Therefore the systemic velocities of the two stars were not forced to be equal and circular orbits were fitted; a small eccentricity does not significantly affect the results. The final spectroscopic orbit is plotted in Fig.~\ref{v615orbitplot} and its parameters are given in Table~\ref{v615v618orbitdata}. The derived minimum masses are $M_{\rm A} \sin^3 i = 4.072 \pm 0.055$\Msun\ and $M_{\rm B} \sin^3 i = 3.177 \pm 0.051$\Msun\ and the mass ratio is $q = 0.7801 \pm 0.0098$.

\subsection{V618 Per}

The analysis of V618\,Per was more complicated as the effective temperatures and rotational velocities of the component stars are more uncertain. For this reason {\sc todcor} was run on combinations of synthetic spectra with $\logg = 4.4$. Template spectra were generated for a wide range of microturbulence velocities, rotational velocities and effective temperatures. The values which gave the lowest residuals of the fit were used as starting values for {\sc sbop}, which was run in single-lined Lehmann Filh\'es mode with an external automatic outlier rejection. The results of these calculations suggest that the best template spectra have effective temperatures of 11000\,K and 8000\,K, rotational velocities of 10\kms, and microturbulence velocities of 2\kms\ and 0\kms\ for primary and secondary respectively. These were used to derive radial velocities with {\sc todcor} and estimates were made of the systematic effects on the orbital semiamplitudes due to template mismatch. For effective temperature, rotational velocity and microturbulence velocity, these amount to 0.2, 0.2 and 0.3\kms\ for the primary star and to 0.1, 0.2 and 0.0\kms\ for the secondary, respectively.

Final radial velocities from {\sc todcor} are given in Table~\ref{v618rvtable} and points rejected from the {\sc sbop} analysis are indicated. These points were rejected as their ($O-C$) values were large compared to the other datapoints. Circular orbits were fitted as {\sc sbop} showed negligable orbital eccentricity; a small eccentricity does not significantly affect the final results. Template mismatch errors were added in quadrature and the final quantities are shown in Table~\ref{v615v618orbitdata}. The derived minimum masses are $M_{\rm A} \sin^3 i = 2.323 \pm 0.031$\Msun\ and $M_{\rm B} \sin^3 i = 1.552 \pm 0.025$\Msun\ and the mass ratio is $q = 0.6682 \pm 0.0087$.

\subsection{Radial velocity of h\,Persei}

All four stars under investigation have a systemic velocity around $-44.2$\kms, consistent with the cluster radial velocities found by Oosterhoff (1937), Bidelmann (1943), Hron (1987), Liu, Janes \& Bania (1989, 1991) and Chen, Hou \& Wang (2003). From the dEBs we can redetermine the radial velocity of h\,Per, using a weighted average over the four stars, to be $44.2 \pm 0.3$\kms. This figure is based on only two stellar systems so the precision of its determination is greater than the accuracy with which it gives the cluster velocity. We conclude that V615\,Per and V618\,Per are almost certainly members of h\,Per.

\begin{figure} \includegraphics[width=0.48\textwidth,angle=0]{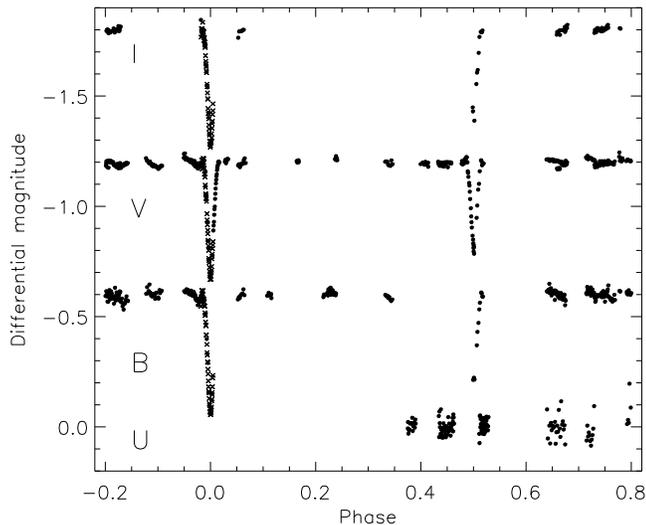} \\
\caption{\label{v615lcs} Phased broad-band filter light curves for V615\,Per. KPK99 data are represented by filled circles and our JKT service data around the primary eclipse is shown using crosses. Light curves in the $B$, $V$ and $I$ filters are offset by $-0.6$, $-1.2$ and $-1.8$\,mag respectively.} \end{figure}


\section{Light curve analysis}      \label{lightcurves}

\subsection{V615 Per}               \label{v615ebop}

V615\,Per is well suited to a photometric analysis using {\sc ebop}\footnote{Eclipsing Binary Orbit Program written by Dr.\ Paul B.\ Etzel (\texttt{http://mintaka.sdsu.edu/faculty/etzel/}).} (Nelson \& Davis 1972, Popper \& Etzel 1981). This is a simple and efficient light curve fitting code where stars are modelled using triaxial ellipsoids. The eclipses of V615\,Per are deep but not total. It is known that in such cases the ratio of the radii of the two stars, $k$, can be relatively poorly constrained (see e.g., Clausen \etal\ 2003), particularly when the component stars are sufficiently well separated to have no discernable reflection effect. This causes $k$ and the ratio of the surface brightnesses of the components to be significantly degenerate. This indeterminacy can be lifted in the case of V615\,Per by our knowledge of a light ratio of the two stars from spectroscopy.

The KPK99 light curves and the JKT service data were combined and phased using the ephemeris derived in section~\ref{v615period} and the resulting data investigated using {\sc ebop}. The $UBVI$ light curves are shown in Figure~\ref{v615lcs}. There is no photometric or spectroscopic indication of extra light from a third star close to V615\,Per, and light curve solutions were consistent with this, so third light was fixed at zero. The secondary eclipse cannot be fitted properly without a small amount of orbital eccentricity, so the quantities $e\cos\omega$ and $e\sin\omega$ were allowed to vary in all solutions, where $e$\,=\,orbital eccentricity and $\omega$\,=\,longitude of periastron of the binary orbit. Filter-specific linear limb darkening coefficients were taken from van Hamme (1993), gravity darkening exponents $\beta_1$ were fixed at 1.0 (Claret 1998) and the mass ratio was fixed at the spectroscopic value.

\begin{figure} \includegraphics[width=0.48\textwidth,angle=0]{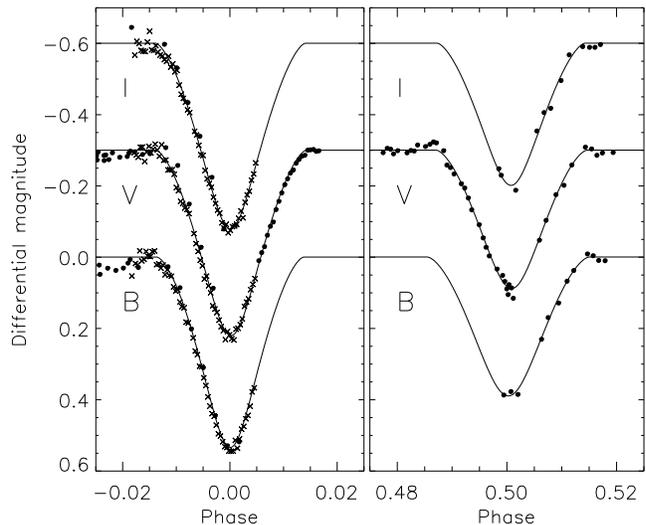} \\
\caption{\label{v615lcfit} Best {\sc ebop} model fits to the light curves of V615\,Per (see Table~\ref{v615lctable}). KPK99 data are represented by filled circles and our JKT service data around the primary eclipse is shown using crosses.
Light curves in the $V$ and $I$ filters are offset by $-0.3$ and $-0.6$\,mag respectively.} \end{figure}

\begin{table*} \begin{center} \caption{\label{v615lctable}Parameters of the light curve fits for V615\,Per using {\sc ebop}. The final values and uncertainties represent confidence intervals. Super- and sub-scripted errors represent the effects of changing the spectroscopic light ratio by its own uncertainty.}
\begin{tabular}{lllll} \hline \hline
\                                               & $B$       & $V$       & $I$       & Adopted value   \\ \hline
Limb darkening coefficient $u_{\rm A}$          & 0.348     & 0.301     & 0.189     &                 \\
Limb darkening coefficient $u_{\rm B}$          & 0.440     & 0.380     & 0.238     &                 \\
Luminosity ratio $\frac{L_{\rm B}}{L_{\rm A}}$        
& $0.512^{+0.071}_{-0.064}$   & $0.555^{+0.077}_{-0.070}$   & $0.594^{+0.082}_{-0.075}$   &                     \\
\hline                        Ratio of the radii ($k$)          
& $0.796^{+0.064}_{-0.056}$   & $0.840^{+0.057}_{-0.058}$   & $0.864^{+0.067}_{-0.059}$   & $0.833 \pm 0.058$   \\
\hline      \vspace*{4pt}     Radius of primary star in units of semimajor axis ($r_{\rm A}$)  
& $0.0509^{-0.0013}_{+0.0013}$& $0.0488^{-0.0013}_{+0.0013}$& $0.0475^{-0.0019}_{+0.0013}$& $0.0491 \pm 0.0030$ \\
            \vspace*{4pt}     Radius of secondary star in units of semimajor axis ($r_{\rm B}$)
& $0.0405^{+0.0013}_{-0.0017}$& $0.0410^{+0.0016}_{-0.0019}$& $0.0410^{+0.0013}_{-0.0017}$& $0.0408 \pm 0.0020$ \\
            \vspace*{4pt}     Central surface brightness ratio ($J$)  
& $0.838^{-0.003}_{+0.012}$   & $0.811^{-0.003}_{+0.006}$   & $0.810^{-0.003}_{+0.004}$   &                     \\
            \vspace*{4pt}     Inclination ($i$) (degrees)           
& $88.83^{-0.10}_{+0.19}$     & $88.76^{-0.07}_{+0.12}$     & $88.81^{-0.05}_{+0.10}$     & $88.80 \pm 0.20$    \\
                              Orbital eccentricity ($e$)
& $0.0396^{+0.0012}_{-0.0011}$& $0.0105^{-0.0022}_{+0.0037}$& $0.0189^{+0.0054}_{+0.0013}$& $0.01 \pm 0.01$     \\
\hline \hline \end{tabular} \end{center} \end{table*}

Solutions were made for many different values of $K$ and for the $BVI$ light curves separately. The residuals of the fit were almost the same for $0.75 < k < 1.1$. The light ratio at 4481\,\AA\ found with spectral synthesis was converted to values for the $BVI$ filters using {\sc ATLAS9} fluxes convolved with filter and CCD efficiency functions\footnote{Filter transmission functions and the quantum efficiency function of the SITe2 CCD used to observe our JKT service data were taken from the Isaac Newton Group website \texttt{http://www.ing.iac.es/Astronomy/astronomy.html}.}. 

Corresponding values of $k$ were derived and used to determine the individual stellar radii, the surface brightness ratio and the orbital inclination. The best {\sc ebop} fits are shown in Fig.~\ref{v615lcfit}. The results for each light curve together with the adopted values are given in Table~\ref{v615lctable}. The upper and lower bounds quoted for individual quantities show the effect of changing the light ratio within the errors quoted. The adopted results include this source of error and a contribution from other error sources, for example the period used to phase the light curves.

It is notable that the primary radius and $k$ (but not the secondary radius), show a systematic variation with filter wavelength. Eclipse depths are known to depend on wavelength when a dEB contains stars of different effective temperatures and therefore colours, but such a variation is not noticable in the current low-quality light curves. This inconsistency should be resolved when better light curves are obtained.

\subsection{V618 Per}               \label{v618ebop}

\begin{figure} \includegraphics[width=0.48\textwidth,angle=0]{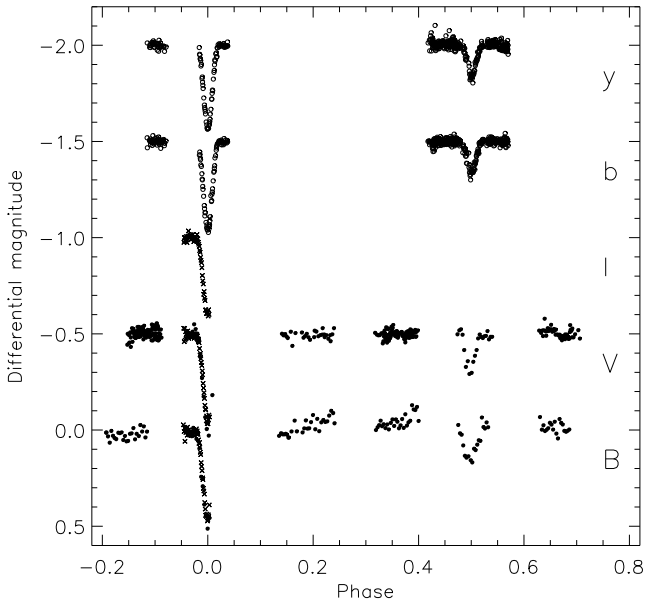} \\
\caption{\label{v618lcs} Phased light curves for V618\,Per. KPK99 data is represented by filled circles and our JKT service data around the primary eclipse is shown using crosses. Str\"omgren data is represented by open circles. light curves in $V$, $I$, and Str\"omgren $b$ and $y$ filters are offset by $-0.5$, $-1.0$, $-1.5$ and $-2.0$\,mag respectively.} \end{figure}

V618\,Per is a more difficult case to analyse using our present light curves. Its spectroscopic light ratio is uncertain. Its eclipses are also less deep than V615\,Per, causing a strong degeneracy between $k$ and the ratio of the surface brightnesses of the two stars. All known light curves are shown in Figure~\ref{v618lcs}. The KPK99 $B$ light curve suggests a slight reflection effect outside eclipse but this is not present in the $V$ light curve. Our Str\"omgren data from 2002 December and 2003 January suffer less from night errors than the data for V615\,Per. In the absence of high quality light curves, we have compensated for the night errors with slight offsets for different nights, and included the data in the analysis with {\sc ebop} along with the $BV$ KPK99 and JKT service light curves. Filter-specific linear limb darkening coefficients were taken from van Hamme (1993), gravity darkening exponents $\beta_1$ were fixed at 1.0 (Claret 1998) and the mass ratio was fixed at the spectroscopic value. Third light was set to zero; there is no photometric or spectroscopic indication of contaminating light. 

\begin{table} \begin{center} \caption{\label{v618lctable} Parameters of the light curve fits for V618\,Per using {\sc ebop}. Parameter designations are as in Table~\ref{v615lctable}. Uncertainties are confidence intervals (see text for 
details).}
\begin{tabular}{llllll} \hline \hline
\                       & $B$       & $V$       & $b$       & $y$       & Adopted               \\ \hline
$u_{\rm A}$             & 0.440     & 0.380     & 0.432     & 0.379     &                       \\
$u_{\rm B}$             & 0.575     & 0.509     & 0.571     & 0.509     &                       \\
$r_{\rm A}$             & 0.0876    & 0.0737    & 0.0733    & 0.0715    & $0.072 \pm 0.003$     \\
$k$                     & 0.831     & 0.816     & 0.804     & 0.800     & $0.802 \pm 0.010$     \\
$r_{\rm B}$             & 0.0728    & 0.0601    & 0.0590    & 0.0572    & $0.058 \pm 0.003$     \\
$J$                     & 0.480     & 0.474     & 0.407     & 0.429     &                       \\
$i$ (\degr)             & 86.4      & 87.4      & 87.1      & 87.1      & $87.1 \pm 0.5$        \\
$e$                     & 0.0646    & 0.0658    & 0.00462   & 0.0119    & $0.01 \pm 0.01$       \\
$\omega$ (\degr)        & 266.9     & 266.6     & 276.6     & 274.2     & $275 \pm 2$           \\
\hline \hline \end{tabular} \end{center} \end{table}

\begin{figure} \includegraphics[width=0.48\textwidth,angle=0]{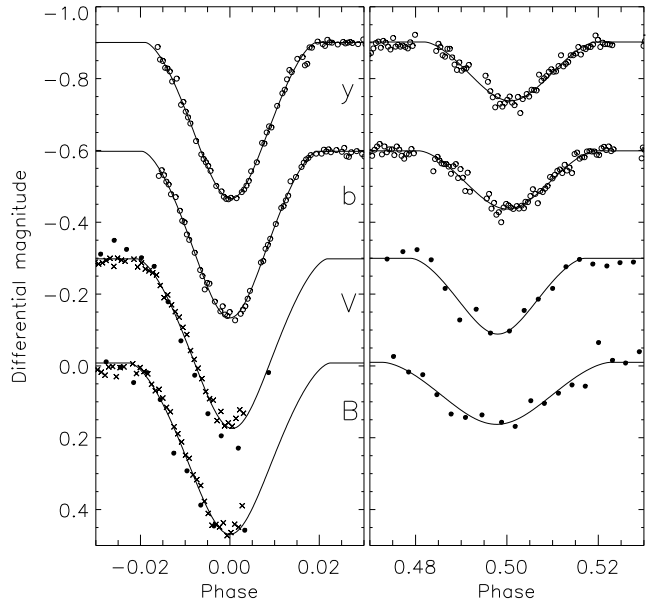} \\
\caption{\label{v618lcfit} Best {\sc ebop} model fits to the light curves of V618\,Per (see Table~\ref{v618lctable}). KPK99 data are represented by filled circles and our JKT service data around the primary eclipse is shown using crosses. Str\"omgren data is represented by open circles. light curves in $V$, $b$ and $y$ filters are offset by $-0.3$, $-0.6$ and $-1.2$\,mag respectively.} \end{figure}

Table~\ref{v618lctable} gives the best fitting parameters for the $BVby$ light curves of V618\,Per using a version of {\sc ebop} modified to use the downhill simplex minimisation algorithm (Press \etal\ 1992) to find the best fit. Figure~\ref{v618lcfit} shows that there is disagreement between the eclipse depths in $B$ and $V$ between the JKT service data (crosses) and the KPK99 light curves (filled circles). Combined with the sparseness of the data during secondary eclipse, this renders the $BV$ light curve solutions unreliable. The adopted photometric parameters of V618\,Per (Table~\ref{v618lctable}) have therefore been taken from the $b$ and $y$ light curve solutions with a significant uncertainty added to account for degeneracy between $k$ and the primary stellar radius. Figure~\ref{v618lcfit} shows the {\sc ebop} model fits to primary and secondary eclipses. 

\section{Absolute dimensions and comparison with stellar models}      \label{dimensionsmodels}

Table~\ref{absolutedimensions} contains absolute dimensions and radiative properties of V615\,Per and V618\,Per calculated from the results of spectroscopic, photometric and spectral synthesis analyses. An important check is whether the surface gravity values of the stars are consistent. Except for the primary component of V615\,Per, they are all close to the expected values for zero age main sequence (ZAMS) stars. V615\,Per\,A is the most massive star being studied here and has a marginally lower surface gravity consistent with slight evolution away from the ZAMS.

\begin{table*} \begin{center} \caption{\label{absolutedimensions}
Absolute dimensions of the detached eclipsing binaries V615\,Per and V618\,Per in the open cluster h\,Persei. \newline
$^*$\,Calculated using the combined system magnitudes in the $V$ filter, light ratios found using the $V$ (V615\,Per) and $y$ (V618\,Per) filter light curves, the assumed cluster distance modulus and reddening and the canonical reddening law $A_V = 3.1\EBV$.
\newline $^\dag$\,Calculated using the effective temperature -- bolometric correction calibration of Strai\u{z}ys \& Kuriliene (1981)}
\begin{tabular}{l r@{\,$\pm$\,}l c r@{\,$\pm$\,}l c r@{\,$\pm$\,}l c r@{\,$\pm$\,}l} \hline \hline
\hspace*{70pt}    & \multicolumn{2}{c}{V615\,Per A} & \hspace*{10pt} & \multicolumn{2}{c}{V615\,Per B} & 
\         \hspace*{20pt} & \multicolumn{2}{c}{V618\,Per A} & \hspace*{10pt} & \multicolumn{2}{c}{V618\,Per B} \\ \hline
Cluster age $\log\tau$ (years)& \multicolumn{11}{c}{7.10 $\pm$ 0.01} \\
Cluster distance modulus      & \multicolumn{11}{c}{11.70 $\pm$ 0.05} \\
Period (days)                 &\multicolumn{5}{c}{13.71390 $\pm$ 0.00002}&&\multicolumn{5}{c}{6.366696 $\pm$ 0.000004}\\
Mass ratio $q$                & \multicolumn{5}{c}{0.7801 $\pm$ 0.0098} & & \multicolumn{5}{c}{0.6682 $\pm$ 0.0087} \\
Mass (\Msun)                  & 4.075   & 0.055 && 3.179   & 0.051 && 2.332   & 0.031 && 1.558   & 0.025 \\
Radius (\Rsun)                & 2.291   & 0.141 && 1.903   & 0.094 && 1.636   & 0.069 && 1.318   & 0.069 \\
Surface gravity \logg         & 4.328   & 0.059 && 4.381   & 0.050 && 4.378   & 0.042 && 4.391   & 0.052 \\ 
Effective temperature (K)     & 15\,000 & 500   && 11\,000 & 500   && 11\,000 & 1000  && 8\,000  & 1000  \\
$M_V$$^*$                     & 0.03    & 0.11  && 0.66    & 0.15  && 1.42    & 0.09  && 2.80    & 0.15  \\
Luminosity log\,($L$/\Lsun)$^\dag$ & 2.37 & 0.08 && 1.82   & 0.10  && 1.51    & 0.14  && 0.77    & 0.08  \\ 
Rotational velocity (\kms)    & 28      & 5     && 8       & 5     && 10      & 5     && 10      & 5     \\
Synchronous \Vrot\ (\kms)     & 8.45 & 0.52 && 7.02 & 0.35 && 13.01 & 0.55 && 10.48 & 0.55 \\
Systemic velocity (\kms)      & $-$44.27 & 0.73 & & $-$44.08 & 0.54 & & $-$44.42 & 0.82 & & $-$44.29 & 0.52 \\
\hline \hline \end{tabular} \end{center} \end{table*}

\subsection{Stellar and orbital rotation}

All four stars of the dEBs are slow rotators compared to nearby single B\,stars (Abt, Levato \& Grosso 2002); only V615\,Per\,A has a measured rotational velocity greater than the synchronous value. The timescales of rotational synchronisation for these two dEBs are $\sim$500\,Myr and $\sim$25\,Myr for V615\,Per and V618\,Per respectively (Zahn 1977, Hilditch 2001). For V618\,Per this timescale is of the same order of magnitude as the cluster age so it is to be expected that the component stars have had their rotational evolution affected by their membership of a close binary system. For V615\,Per, however, the timescale is much greater so the rotational velocities of the stars will not have changed significantly during their main sequence lifetime. Therefore their slow rotation must be primordial in nature (see Valtonen 1998). 

The timescales of orbital circularisation for the two dEBs are $\sim$1200\,Gyr and $\sim$21\,Gyr for V615\,Per and V618\,Per respectively (Zahn 1977, Hilditch 2001) so it is expected that these young systems will have undergone no orbital evolution during their main sequence lifetimes. Both dEBs exhibit negligible eccentricity and their closeness to circularisation must again be primordial (Zahn \& Bouchet 1989).

\subsection{Stellar model fits}

The physical parameters of the four stars in V615\,Per and V618\,Per have been compared to two different sets of stellar models, the Granada models (Claret 1995, 1997; Claret \& Gim\'enez 1995, 1998) and the Padova models (Girardi \etal\ 2000). The Granada models are available for four metallicities, each with three different hydrogen abundances: $Z = 0.004$ ($X = 0.65, 0.744, 0.80$), $Z = 0.01$ ($X = 0.63, 0.73, 0.80$), $Z = 0.02$ ($X = 0.60, 0.70, 0.80$) and $Z = 0.03$ ($X = 0.55, 0.65, 0.75$). Models for stars with masses below 1.2\Msun\ have no overshooting; more massive model stars have moderate overshooting. The Padova models are available for six metallicities: $Z = 0.0004, 0.001, 0.004, 0.008, 0.019, 0.030$, with one helium abundance for each. Models at all metallicities have moderate overshooting and the $Z = 0.019$ models (solar metallicity) are also available with no overshooting.

\begin{figure*} \includegraphics[width=\textwidth,angle=0]{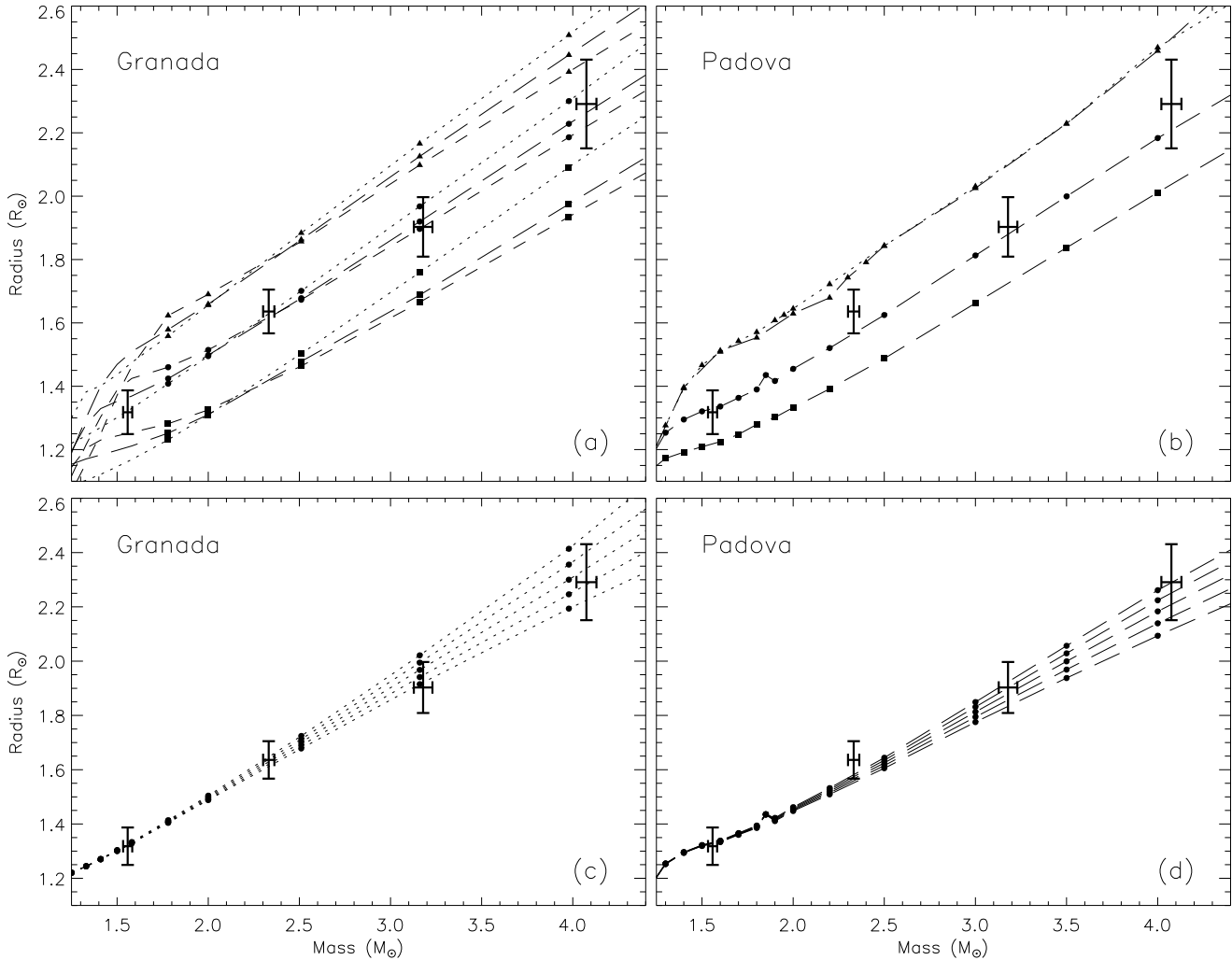} \\
\caption{\label{modelMRfit} Comparison of stellar evolutionary models to the masses and radii of the stars of V615\,Per 
and V618\,Per for two different sets of theoretical models. (a) Granada models plotted for metallicities $Z = 0.004$ 
(squares), $Z = 0.01$ (circles) and $Z = 0.02$ (triangles). Each metallicity is available in three hydrogen abundances 
(see text) which are plotted with long dashes for standard, short dashes for lower, and dots for higher hydrogen 
abundance. For clarity, symbols are shown only for masses above 1.75\Msun. (b) Padova models plotted for metallicities 
(bottom to top) $Z = 0.004$, $Z = 0.008$, and $Z = 0.019$. The $Z = 0.019$ track with no convective overshooting is 
shown using a dotted line. (c) Granada models with ($X$, $Z$) = (0.63, 0.01) plotted for ages (bottom to top) of 3, 8, 
13, 18 and 23 Myr. (d) Padova models with ($Y$, $Z$) = (0.25, 0.008) plotted for ages (bottom to top) of 3, 8, 13, 18 
and 23 Myr.} \end{figure*}

The two sets of models have been plotted in the mass--radius plane, with the two dEBs, for three metallicities and for an age of 13\,Myr ($\log\tau = 7.11$) in Fig.~\ref{modelMRfit}(a)(b). A best fit is obtained using the Granada models with ($X$, $Z$) = (0.63, 0.01), although panel (b) suggests the Padova models would fit equally well for the same $Z = 0.01$. Panels (c) and (d) of Fig.~\ref{modelMRfit} show the best-fitting evolutionary models for the Granada and Padova sets, for ages of $\log\tau = 6.47, 6.90, 7.11, 7.26, 7.36$ (years). 

\section{Discussion}    \label{discussion}

We have derived absolute dimensions for two early-type detached eclipsing binaries in the young open cluster h\,Per. Spectral synthesis has given the effective temperatures and rotational velocities of both systems. The negligibly eccentric orbits and low rotational velocities of all four stars supports the `delayed break-up' route of binary star formation (Tohline 2002). In this scenario a protostellar core embedded in a molecular cloud contracts towards the ZAMS. It accretes material with a high specific angular momentum from the surrounding cloud, and spins up whilst losing gravitational potential energy. When the ratio of rotational energy to the absolute value of gravitational energy, $\beta$, reaches approximately 0.27 (Lebovitz 1974, 1984), the core deforms into an ellipsoidal shape. From this it forms a `dumbbell' shape which splits to form a binary system with a circular orbit and low stellar rotational velocities. Other examples exist of young long-period spectroscopic binaries with circular orbits, e.g., \#363 in NGC\,3532 (Gonz\'alez \& Lapasset 2002).

The four stars exhibit a good spread of masses and radii which should provide an excellent test of stellar evolutionary models. The radii could not be determined from the current light curves with great accuracy, so the analysis has been restricted to a determination of the bulk metallicity of the h\,Per cluster of $Z \approx 0.01$. The existence of Galactic disc low-metallicity young B\,stars is already known (e.g., GG\,Lupi; Andersen, Clausen \& Gim\'enez 1993).  

The chemical composition of h and $\chi$ Per has been investigated many times with somewhat conflicting results. Nissen (1976) found the helium abundance of h\,Per to be significantly lower than that of field stars, based on narrow-band photometry of twelve ZAMS and slightly evolved B\,stars. This conclusion was supported by the spectroscopic observations of Wolff \& Heasley (1985). However, from high-resolution spectroscopic abundance analyses of four stars, Lennon, Brown \& Dufton (1988) and Dufton \etal\ (1990) found that helium abundance was normal and suggested that the surface gravities derived by Nissen (1976) were too low. Dufton \etal\ also found that h and $\chi$ Per have approximately solar metal abundances. This conclusion was supported by Smartt \& Rolleston (1997), but Vrancken \etal\ (2000) find that the abundances of various metals are 0.3--0.5\,dex below solar from abundance analyses of eight early B\,type giants.

The above results refer to empirical determinations of the mean photospheric abundances of helium and several light metals. Our derivation of the cluster metallicity, $Z \approx 0.01$, has been found by comparison with theoretical stellar evolutionary models and refers to the overall metal abundance in the interiors of the stars analysed. This quantity is directly relevant to the fitting of theoretical isochrones to the positions of stars in observed CMDs of the h and $\chi$ Per open clusters. 

The four recent photometric studies of h and $\chi$ Persei have not included the effects of non-solar metallicity in their analyses; whilst Marco \& Bernabeu (2001) and Slesnick \etal\ (2002) assumed a metallicity of $Z = 0.02$, the works of Capilla \& Fabregat (2002) and Keller \etal\ (2001) make no mention of metallicity. If all analyses used a solar metallicity, the derived age and distance modulus of h and $\chi$ Per could be systematically incorrect. This possibility also is increased by the dependence on one set of model isochrones; three of the four works used the stellar models of the Geneva Group (Schaller \etal\ 1992), although Keller \etal\ used the previous generation of Padova models (Bressan \etal\ 1993; Bertelli \etal\ 1994). Once more accurate radii for the four stars studied here are obtained, a reanalysis of h and $\chi$ Persei should be undertaken to ensure that reliable parameters are known.

The four recent photometric analyses suggest the age of the cluster is $\log\tau = 7.10 \pm 0.05$ (see section~\ref{clusterinfo}) and the positions of the stars of V615\,Per and V618\,Per in the mass--radius plane are consistent with this. The traditional degeneracy between age, metal abundance and helium abundance (Thompson \etal\ 2001) could be broken with better light curves for the two dEBs. In that case the large range of masses of the four stars would allow the less massive stars to set the metallicity and the more massive stars to set the age of the cluster, with information on the helium abundance contained in the slope of the observational line in the mass--radius plane. Such an analysis would benefit from better sampled grids of stellar models and a greater choice of helium abundance and degree of overshooting (see also Young \etal\ 2001). 

Definitive light curves of the dEBs should give radii to 1--2\% and the individual brightnesses of the component stars in the observed filters. This will allow more discriminate testing of stellar evolutionary models and the construction of a cluster HR diagram with mass and radius determinations for four individual stars. If the light curves are observed in the Str\"omgren $uvby$ or the broad-band $VRK$ filters, the surface brightnesses of the stars could be accurately derived (Moon 1984, di\,Benedetto 1998, Lacy 1978, Barnes, Evans \& Parsons 1976). When combined with the radius determinations of the stars, the distances of each star could be found individually, giving four separate measures of the distance to h\,Per, independent of stellar models and main sequence fitting.

The richness of h and $\chi$ Per suggests there may be undiscovered eclipsing systems which could be added to this analysis. V621\,Per is a member of $\chi$\,Per with a spectral type of B2\,III and is known to be a dEB (Krzes\'\i nski \& Pigulski, 1997). It is a spectroscopic binary but spectra taken during our INT observing run show no sign of a fainter component, making it impossible to find the absolute masses and radii of its component stars in the usual way.


\section*{Acknowledgements}

The authors would like to thank Jens Viggo Clausen for providing {\sc sbop} and {\sc ebop}, a modified version of {\sc ebop} and other resources. We would like to thank the referee for their prompt reply and comments which significantly increased the quality of the paper. We would also like to thank Stephen Smartt for discussions and Liza van Zyl for reading the manuscript.

JS acknowledges financial support from PPARC in the form of a postgraduate studentship. The authors acknowledge the data analysis facilities provided by the Starlink Project which is run by CCLRC on behalf of PPARC. In addition, the following Starlink packages have been used: {\sc convert}, {\sc kappa}, {\sc figaro}, {\sc autophotom} and {\sc slalib}. 
The following internet-based resources were used in research for this paper: the WEBDA open cluster database; the ESO 
Digitized Sky Survey; the NASA Astrophysics Data System; the SIMBAD database operated at CDS, Strasbourg, France; the 
VizieR service operated at CDS, Strasbourg, France; and the ar$\chi$iv scientific paper preprint service operated by Cornell University.


\end{document}